# Phase transformation on heating of an aged cement paste

E.T. Stepkowska [a], J.M. Blanes [b], F. Franco [c], C. Real [b], J.L. Pérez-Rodr´ıguez [b,*]

[a] *Institute of Hydroengineering PAS, 80 953 Gdansk-Oliwa, Poland*
[b] *Instituto de Ciencia de Materiales de Sevilla, Universidad de Sevilla-C.S.I.C., 41092, Spain*
[c] *Universidad de Málaga, 29071 Málaga, Spain*

**Abstract**

The standard cement paste (C-43-St) was studied previously by static heating, SH, immediately after 1 month hydration at $w/c$ = 0.4 [J. Therm. Anal. Calorim. 69 (2002) 187]. This paste after 5-year ageing (unprotected from contact with air) was subject to thermal analysis in air and in argon (DTA, DTG and TG), to XRD at various temperatures, T, in a high temperature chamber, to mass spectroscopy (MS) and to IR spectroscopy. The aim of this study was to compare the results of SH (fresh paste) and of TG (the aged one), to verify the assumptions made on SH interpretation and to check the change in hydration products with ageing as measured by phase transformation on heating ($\Delta M$ versus the final mass). The sorbed water (EV), escaping at 110 °C from the fresh paste, was bound on ageing with a higher energy and escaped did not change in the less compact one C-43-I. C-S-H gel transformed on heating above 600 °C into $C_2S$ and $C_3S$. Portlandite content did at higher temperatures. The joint water content of hydrates and of C-S-H gel increased on ageing by 1–2% in the dense paste C-43-St and 600 °C. Carbon dioxide and/or carbonate ions to form carbonates, were sorbed during ageing and were present in the aged paste in some form not change on ageing. In the air atmosphere it became partly carbonated, which was accompanied by an increase in mass between 500 and undetectable by XRD (amorphous or crypto-crystalline). Sensitivity to carbonation $\Delta M$(700–800 °C) increased highly with ageing.

*Keywords:* Cement hydration; Cement ageing; Adsorbed water; C-S-H gel; Portlandite; Carbonates; thermal analysis; XRD; IR; Thermal analysis

## 1. Introduction

In the previous papers [1,2] the hydration products were estimated from the mass loss, $\Delta M$, on static heating, SH. These hydration products were:

(i) the sorbed water, evaporating at 110 °C as EV = $\Delta M$(110 °C);

(ii) water of hydrates, $\Delta M$(110–220 °C) (compare [3]); morite structure [4], $\Delta M$(220–400 °C), described as

(iii) water in C-S-H gel of the jennite-like and/or tober- the chemically bound water 1.5–5% [3]; in DTA their

peak is observed at 200–300 °C and another one about 400 °C [5];

hydration, $P(H_2O)$ = $\Delta M$(400–800 °C); this assumes no carbonation below 400 °C; (iv) water released from portlandite, which was formed on

(v) sensitivity to carbonation $\Delta M$(600–800 °C), i.e. the which was formed above 400 °C of portlandite re- quantity of $CO_2$ released from calcium carbonate, 600–700 °C [8]. In freshly hydrated paste the carbon acting with atmospheric air [6,7] and decomposed at dioxide is sorbed from air for this reaction.

Portlandite looses water between 400 and 500 °C (see below and [9]), but if $CO_2$ is available, above 400 °C it may form the calcium carbonate [6]. Thus the assumption of $P(H_2O)$ = $\Delta M$(400–600 °C) would not take into account this reaction excluding the part of portlandite formed on hy-
ature range. If some carbonation occurs below 400 °C, an dration and transformed into carbonate within this temper-
overestimation would result, when assuming (iv): the molec- ular mass of $CO_2$ is 44, whereas that of water is 18. This error is not important though in the freshly hydrated paste, containing little or no sorbed $CO_2$, but it would be high in an aged one (see below).

Pastes of some cements and some powders, hydrated in water vapour, were studied previously by static heating (SH)

* Corresponding author. Fax: +34 95 4460665.
*E-mail address:* jlperez@cica.es (J.L. Pe´rez-Rodr´ıguez).

right after termination of the hydration process [1,2]. These pastes and powders were stored for 5 years, unprotected from contact with air. They were tested by thermal analysis (DTA, DTG and TG) showing the influence of certain parameters. Comparison with respective values measured on freshly hydrated mortars indicates the contribution of ageing. Some additional tests were done on selected samples by other methods (XRD, mass spectroscopy, MS, and IR spectroscopy).

Here only the results obtained for one cement will be presented: the aim of this paper is to check the change in hydration products on ageing. Thus the phase transformation on heating of the freshly hydrated paste (studied in SH), is compared with the aged one (studied by TG). This permits also the verification of the assumptions made to interpret the results of SH.

Experimental

Standard cube (C-43-St) was prepared, at w/c = 0.4, from an Indian OPC of grade 43 [IS: 8112-1989, NCB, New Delhi) described in [1,2] and also the details of the methods are presented there.

The chemical composition was: CaO 61.0%, SiO2 20.9%, Al2O3 5.3%, Fe2O3 3.1%, MgO 3.6%, K2O 0.89%, SO3 1.5%, Na2O 0.45%, LOI 2.7%.

by hand at distilled water/cement ratio, w/c = 0.4. Both Another cube of a lower density (C-43-I) was compacted pastes were hydrated for 1 month at room temperature, air 110 °C overnight, 220 °C for 8 h, 400 °C for 4 h, 600 °C for 2 h and 800 °C for 1 h. The mass loss, $\Delta M$, was calculated dried and studied by static heating, SH (in triplicate) at versus the final mass at 800 °C.

These pastes were aged for 5 years, unprotected from contact with air, and studied by:
thermal analysis, DTA, DTG and TG, at atmospheric 1 K min−1, up to 1000 °C); pressure, either in air or in argon (Seiko TG/DTA 6300,
XRD at 5 K min−1, at selected temperatures of the peaks on DTG curves, i.e. 30, 230, 300, 319, 349, 415, 445, 470, 490, 600, 720 and 760 °C (Philips X-pert-Pro, with high temperature chamber ANTON PARR, HTK 1200, X'celerator, copper tube and θ–θ goniometer);
mass spectrometry, MS of H2O and CO2 in C-43-I (of tronic microbalance at the heating rate of 5 K min−1 from room temperature to 800 °C, under vacuum 4 × the highest total mass loss), measured on a CI elec- 10−5 mbar. The sample tube was attached to the Balzers Quadstar quadrupole QMS200 mass spectrometer
and the H2O and CO2 were determined of molecular mass 18 and 44, respectively;
temperatures, T, i.e. room T, 110, 220, 400 and 600 °C, (iv) IR spectroscopy in KBr disc after heating at various 24 h heating in each case (NICOLETTE, FT-IR 510).

The values of mass loss at the given temperature, $\Delta M(T)$, sample mass at 800 (SH) or 1000 °C (TG). The standard deviation (SH) was ±0.02–0.3%. were calculated in weight percent in relation to the final

## 2. Results and discussion

### 2.1. Thermal behaviour

Results of (i) static heating (SH) of freshly hydrated paste and (ii) of the thermo-gravimetry (TG) of the aged one, C-43-St (mass loss $\Delta M$ in wt.% versus the final mass) are shown in Fig. 1. Respective values, obtained on C-43-St and C-43-I, are interpreted in Table 1.

The results of static heating are comparable with the straight linear sections on the TG curve, in agreement with the introduction (Fig. 1, see also Fig. 6a and c, where the results of thermal analysis and mass spectroscopy, MS, are compared). TG and DTG curves are more precise tools than DTA curves for identification and quantification of all the main hydrated phases [10]:

(a) An extremum at 100–130 °C (TG) corresponds to the escape of sorbed water, EV, at 110 °C (SH).
(b) Between 110 and 220 °C (SH) and between 160 and 185 °C (DTG) the mass loss occurs, defined here as hydrate water.

DTG curve between 220 and 400 °C at about 240, 280 and 380 °C (jennite-like phase [3–5]; this region was (c) Two or three low and broad maxima are observed on the defined in SH study as gel water (220–400 °C).

and DTG curves at 450–470 °C (TG), representing the decomposition of portlandite, $Ca(OH)_2 \rightarrow CaO$ (d) A narrow and high extremum occurs on both the DA + $H_2O$ [9,10]; the $\Delta M$(400–600 °C), measured by SH on freshly hydrated pastes, resulted in a similar value.

(e) Between 500 and 700 °C (TG) a further dehydration and/or dehydroxylation and formation of calcite occurs. Thus simultaneously the mass loss (increasing with hydration time [10]) and the mass gain takes place and the resultant is measured. This range was not analyzed in SH.

(f) The decomposition of calcite occurs at 680–770 °C (TG) [6,8]; the $\Delta M$(600–800 °C), representing this process, was defined in SH as sensitivity to carbonation.

Thus the straight linear sections on the TG curves correspond generally to the temperature ranges mentioned above, covering the following: between 60–90 and 150, 150–400, 400–500, a complex region 500–700 and finally 680–770 °C.

When comparing the mass loss in the fresh paste (SH) served below 150 °C and above 500 °C (Fig. 1). Between 150 and 400–500 °C the TG curves are little dependent on with the aged one (TG), the main difference could be observed the measurement conditions (in air or in argon) and they run almost parallel to the SH line. The slope $\Delta M/\Delta T$, i.e. the

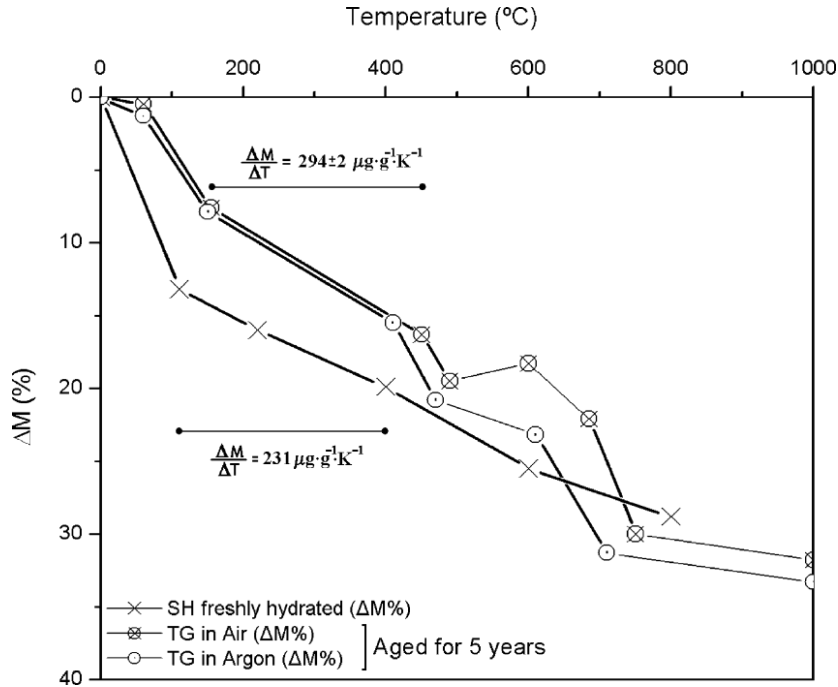

Fig. 1. Results of (i) static heating (SH) of freshly hydrated paste and (ii) of the thermo-gravimetry (TG) of the aged one, C-43-St (mass loss $\Delta M$ in weight % vs. the final mass).

similar, i.e. 294 and 231, respectively (in $\mu g\ g^{-1}\ K^{-1}$). rate of the mass change with temperature is in both cases

A much higher mass loss in SH than in TG was measured up to about 400 °C. Thus a part of the sorbed water (EV) higher temperature to escape, i.e. 400–500 °C (instead of 110 °C). At this temperature the mass loss in SH and TG (in was bound with time with a higher energy and needed a argon) is similar, whereas it is somewhat smaller in TG if measured in air. In this case the mass increased on further heating; apparently $CO_2$ was sorbed from the surrounding air. At about 700 °C the difference in $\Delta M$ disappeared but the total mass loss up to 800 °C was higher in TG than in SH (Fig. 1).

The thermal reactions were generally endothermic (DTA 220 and about 400 °C an exothermic reaction occurred in and DTG, Fig. 6a and c). Only within the range between the aged pastes studied in air (Fig. 6c). This effect was not observed in argon atmosphere (Fig. 6a): it may be attributed to the combustion of the organic matter formed during the 5-year ageing of the hydrated pastes.

Thus important changes in the mass loss on heating occurred due to ageing and the results of the thermal study may be summarized as follows:

The escape of sorbed water (up to 110 °C in SH) moved with ageing to higher temperatures; only about up to 150 °C from the aged specimens. one-half (C-43-St), or one-third (C-43-I) evaporated

The increase in C-S-H gel water content in aged paste is indicated by the DTG peaks between 200 and 400 °C was not important (1–2% only); the formation of jennite [5], see Fig. 6c.

aged pastes (TG) between 500 and 700 °C, possi- (iii) The rest of the sorbed water (SH) escaped from ($9CaO\cdot 6SiO_2\cdot 11H_2O$ [11]). A mass increase occurred bly on dehydroxylation of the jennite-like compound within this temperature range in aged pastes (TG in air) due to $CO_2$ absorption.

Water escaping on portlandite decomposition was the highest in freshly hydrated paste (SH), it was only slightly smaller in aged paste (TG in argon) and it was the lowest in TG in air due to the carbonation on heating (C-43-St). These values were smaller in the less compact paste C-43-I, but the quantitative relations were similar.

Table 1
Mass loss $\Delta M$ in weight % vs. the final mass

| Paste | Test | Sorbed water | C-S-H gel and hydrates | Portlandite water | High $T$ water | $CO_2$ carbonate | Mass increase |
|---|---|---|---|---|---|---|---|
| C-43-St | $T$ in SH | 110° C | 110–400° C | 400–600° C | – | 600–800° C | – |
| Fresh SH | | 13.2 | 6.7 | 5.6 | – | 3.3 | – |
| Aged TG air | | 7.6 | 8.7 | 3.2 | 2.6* | 7.9 | 1.2* |
| Aged TG argon | | 7.9 | 7.6 | 5.3 | 2.4 | 8.1 | – |
| C-43-I | $T$ in TG | 170° C | 150–450° C | 450–500° C | 500–700° C | 700–800° C | |
| Fresh SH | | 17.3 | 5.6 | 4.4 | – | 3.4 | – |
| Aged TG air | | 6.5 | 5.5 | 2.5 | 10.9 | 14.5 | – |
| Aged TG argon | | 6.8 | 4.5 | 3.4 | 13.3 | 12.0 | – |

(iv) On calcite decomposition (600–800 °C), more $CO_2$ evolved from the aged paste (TG); it was similar in air and argon atmosphere (standard paste, C-43-St). Thus $CO_2$ was absorbed mainly during ageing, and to lesser extent—during heating in air. Also the total mass loss was higher in the aged paste, but it was again slightly higher in argon than in air, in spite of its mass increase on heating up to 600 °C. No influence of paste density was observed in the freshly hydrated pastes, whereas much more $CO_2$ escaped from the paste C-43-I, than from the standard one (Table 1).

(v) Calcite may possibly form not only on portlandite carbonation but also from other compounds, like amorphous carbonate or C-S-H gel.

(i) The paste of a lower density indicated a much higher sorbed water content (SH) and after ageing (TG)—an important content of high temperature water and an elevated sensitivity to carbonation (TG), in this case higher in air than in argon (C-43-I, see table in Fig. 1).

To explain the observations concerning carbonates, additional tests were done by XRD, IR and MS.

### 2.2. X-ray diffraction characterization

XRD of the aged paste C-43-St (at selected temperatures of the peaks on DTG curves) are shown in Fig. 2. The hydrated paste C-43-St, stored for 5 years indicated at room temperature the XRD peaks of the following compounds:
Portlandite of the highest intensity: (0 0 1) at 4.927 Å and 2θ—18.0°, (1 0 0) at 3.111 Å and 28.6°, (1 0 1) at 2.629 Å and 34.1°, (1 0 2) at 1.927 Å and 47.1°, (1 1 0) at 1.795 Å and 50.8°, (1 1 1) at 1.685 Å and 54.4°.
Limited amount of calcite: 3.034 Å at 2θ—29.4°, 2.287 Å at 39.4°, 2.094 Å at 43.2°, joint with aragonite, etc.
Some vaterite: 3.576 Å at 2θ—24.9°, 3.302 Å at 27.0°, 2.747 Å at 32.6° (joint with C3S), 1.821 Å at 50.1°, etc.
Some aragonite: 3.401 Å at 2θ—26.2°, 2.113 Å at 42.8 joint with calcite, etc.
Some unhydrated C2S (Ca2SiO4) and C3S (Ca3SiO5), 2θ—32.2°, 2.755 Å at 32.5°, etc. of content increasing with temperature: 2.780 Å at
A broad band of C-S-H gel between 32° and 33°, i.e. the values of d = 2.74, 2.84, 3.19 Å (9–454) and for 2.80–2.71 Å. The standard files indicate for C-S-H gel tobermorite −2.83 and 11.31 Å (10–373) [3].

To check the phase transformations observed in thermal analysis, the X-ray diffraction study was done at temperatures, at which some thermal effects were observed on DTA and DTG curves, i.e. at the start of the given process, at its peak* and at its termination. These were: room temperature, 230, 300*, 319, 349, 385*, 415, 445, 470*, 490, 600, 630*, 690, 720* and 760 °C. Selected diffractograms are presented in Fig. 2.

Up to 445 °C, the highest XRD peak intensity is indicated by portlandite at 2θ = 18°, i.e. d(00 1) = 4.927 Å. At 230 °C it is slightly lower than at 300 °C and it starts to decrease considerably at 445 °C, disappearing at 470 °C (Fig. 3), in accordance with the DTG results (Fig. 6c).

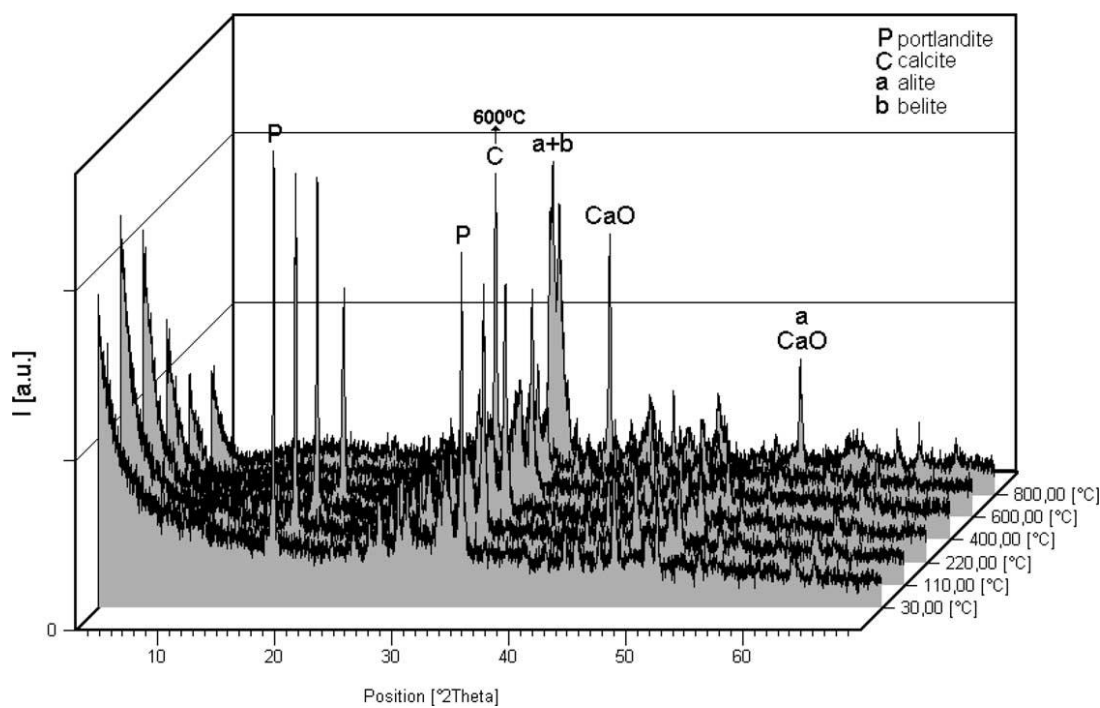

Fig. 2. XRD of the aged paste C-43-St (at selected temperatures of the peaks on DTG curves).

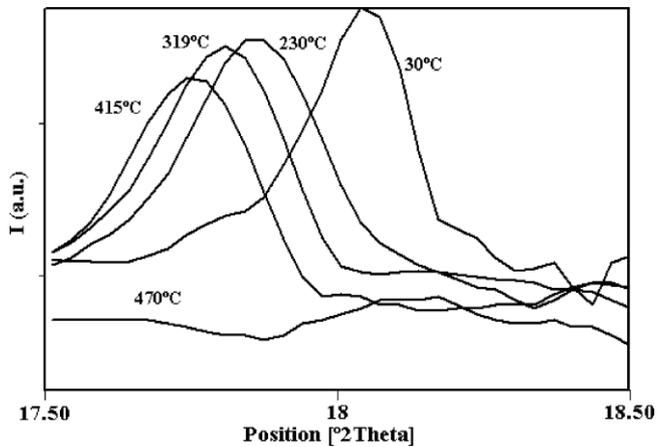

Fig. 3. Portlandite: XRD peak at various temperatures: $d(00\,1)$ = 4.92 Å.

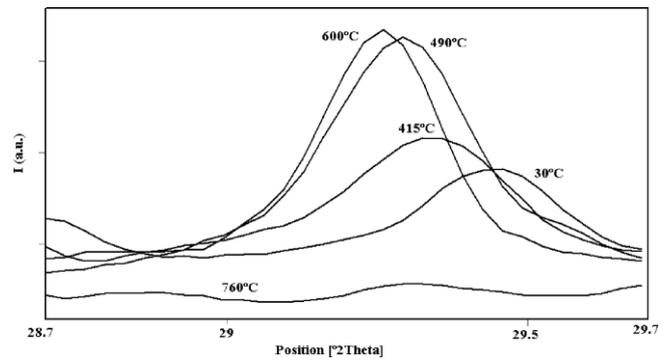

Fig. 4. Calcite: XRD peak at various temperatures: $d$ = 3.036 Å.

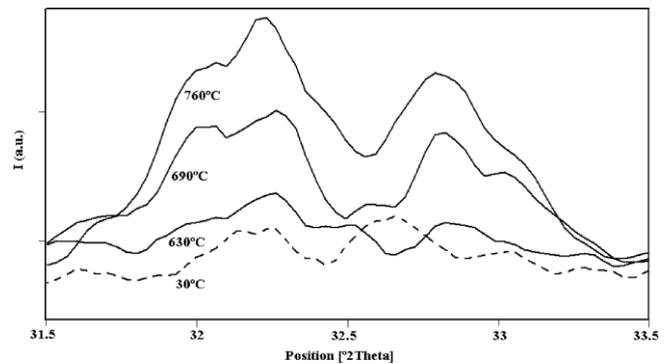

Fig. 5. C-S-H gel: broad band, transforming above 630 °C into $C_2S$ and $C_3S$.

The increase in basal spacing with temperature was observed in portlandite and it was accompanied by the gradual lowering of the peak intensity (Table 2 and Fig. 3). The change in (1 0 1) spacing was much smaller and the d(1 0 0),
of temperature, the peak intensity decreasing up to 445 °C, whereas the (1 0 2) peak was still observable at 470 °C, d(1 0 2), d(1 1 0) and d(1 1 1) were practically independent Table 2.

Main calcite peak at d = 3.036 Å indicated a small in- 445 °C (Fig. 4). On portlandite decomposition at 470 °C tensity at low temperatures, increasing gradually up to the calcite peak increased pronouncedly, it was lowered at 690 °C and disappeared at 720 °C, in accordance with the DTG and TG test results. At 690 °C the CaO peaks were found.

The main basal spacing at 2θ = 29.43°, d = 3.036 Å, It moved to 3.046 Å at 445 °C, and to 3.053 Å at 470 °C, increasing in intensity up to 630 °C, lowering at d = 3.058 Å was less dependent on temperature than that of portlandite. and 690 °C and disappearing at 720 °C. The double peak of aragonite and vaterite at 27.3° and 27.1°, respectively (d
= 3.27 and 3.29 Å) was slightly visible below 400 °C. In the previous study by TEM [1,2], the presence of aragonite and vaterite in this hydrated cement was proved.

The broad band between 32° and 33° of C-S-H gel [3,11] also higher (Fig. 5). At 630 °C within this band a CaO peak was formed at 32.22°, i.e. d = 2.770 Å. At higher temperatures (690 and 720 °C) the formation of CaO increased and considerable amounts of C2S and C3S were observed at d = 2.80 Å and at 2.73–2.78 Å (2θ = 32 and 32.2–32.8°). This was present at room temperature and with its increase, it was water between 500 and 700 °C (TG). A high peak of CaO was also found at 37.12° and 37.06°, i.e. d = 2.422 and was in agreement with the evolution of "dehydroxylation" 2.426 Å, at 720 and 690 °C, respectively.

Table 2
Change with temperature of some spacing [Å] of portlandite

| T (°C) | d(00 1) | I(a.u.) | d(10 0) | d(10 1) | I(a.u.) | d(10 2) | d(11 0) | d(11 1) |
|---|---|---|---|---|---|---|---|---|
| RT | 4.927 | 379 | 3.111 | 2.629 | 300 | 1.927 | 1.795 | 1.685 |
| 230 | 4.955 | 343 | | 2.637 | 268 | | 1.801 | |
| 300 | 4.956 | 333 | 3.111 | 2.640 | 264 | | 1.801 | 1.685 |
| 319 | 4.957 | 333 | | 2.641 | 241 | | 1.801 | |
| 349 | | | | 2.644 | 241 | 1.927 | 1.803 | 1.685 |
| 385 | 4.986 | 328 | 3.121 | 2.644 | 223 | | 1.803 | 1.685 |
| 415 | 4.986 | 313 | 3.121 | 2.652 | 241 | 1.927 | 1.802 | 1.685 |
| 445 | 5.010 | 262 | – | 2.653 | 164 | 1.927 | | 1.685 |
| 470 | – | – | | | | 1.927 | | ? |
| 2θ (°)[a] | 18.04 | | 28.7 | 34.1 | | 47.1 | 50.7 | 54.4 |

[a] [22].

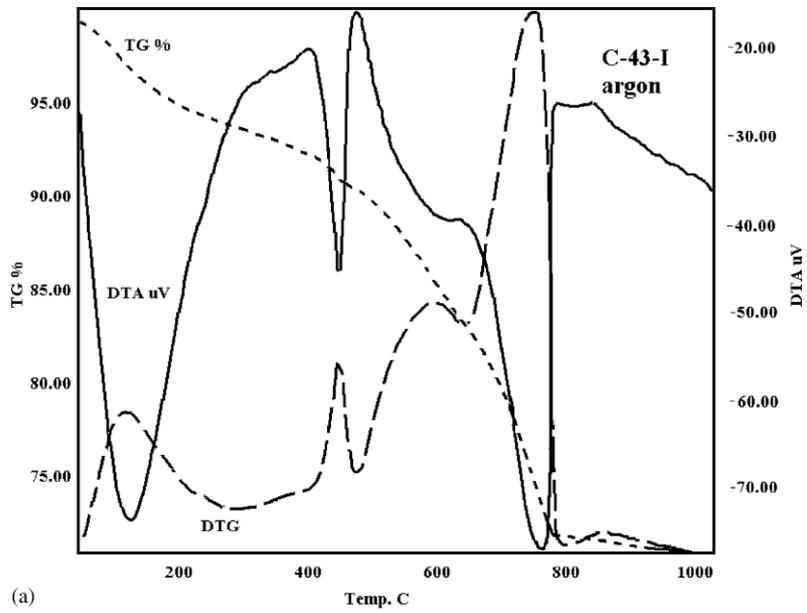
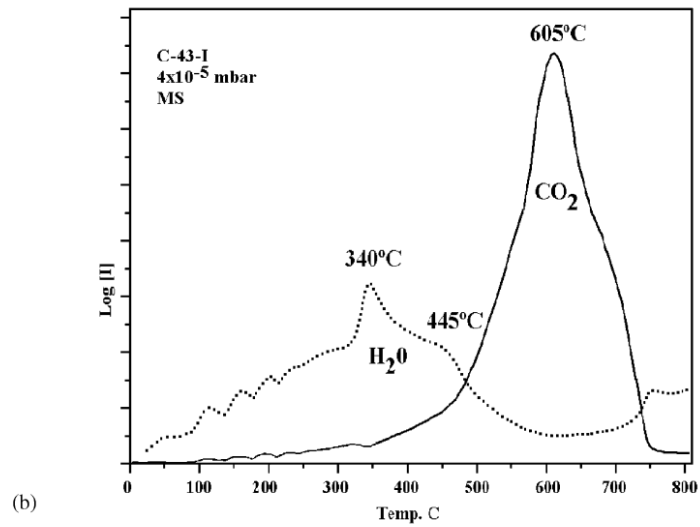
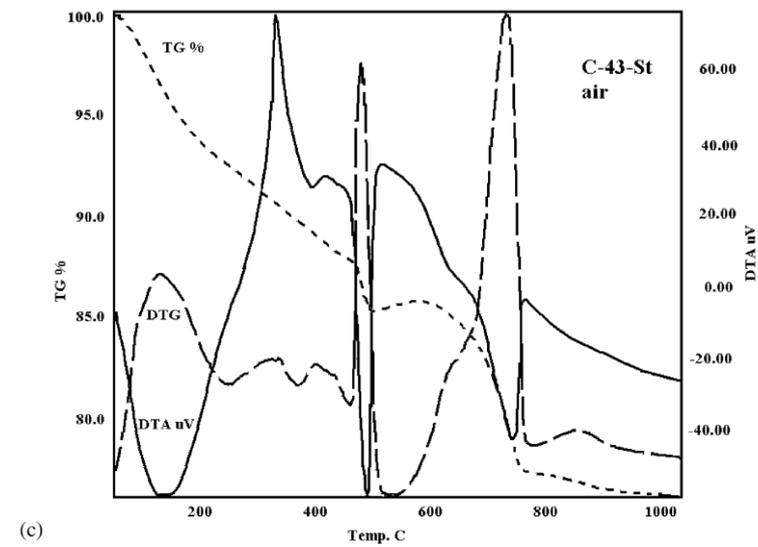

Fig. 6. Comparison of thermal analysis (DTA, DTG and TG) and mass spectrometry results.

470 °C as the broad peak at 2θ = 37.23° and d = 2.415 Å. Its intensity increased considerably at 720 °C. Also its peak The start of CaO formation was observed at 445 and at 53.9°, d = 1.701 Å was found at temperatures exceed- ing 445 °C, after which (445–630 °C) it constituted a broad band, starting at 53.85°, 1.703 Å. At 690–720 °C, a double peak could be found at 53.46° and 53.62°, i.e. d = 1.709 and 1.714 Å.

Thus the assumptions, made on interpretation of the re- sults of static heating of the fresh paste proved to be correct:

Portlandite decomposed above 400 °C, the possible calcite formation occurred around 500 °C and its decomposition above 600 °C.

Mass spectroscopy as compared with thermal analysis

1000 °C), was 39.9%, both as measured by mass spectrome- The total mass loss, related to the final mass (at 800 or try, MS and by thermogravimetry, TG, respectively (Fig. 6).

The mass spectra (Fig. 6b) were obtained in vacuum, thus the peaks representing the escape of the respec- tive compound were observed at lower temperatures than those found in the thermal analysis done at atmospheric pressure. Between 200 and 400 °C two peaks were observed on the DTG curve, characteristic of jennite [4] (Fig. 6c), and perature (Fig. 6b). The peak at 340 °C (MS, Fig. 6b), cor- similar peaks were found on the MS curve at a lower tem- 450 °C (Fig. 6a). A shoulder at 445 °C (MS, Fig. 6b) in- responds to portlandite decomposition, observed in DTG at dicates either the dehydration or dehydroxylation of some compounds formed on ageing and heating, or the escape of water from hexagonal hydrates, or of another form of wa- ter, e.g. of amorphous calcium carbonate hydrate. A gradual decrease in water escape rate follows.

imum at 605 °C (MS, Fig. 6b), corresponding to the 760 °C peak on the DTG curve (Fig. 6a). Between 400 and 500 °C The high and broad peak of $CO_2$ escape, indicates a max- sponding region in DTG is 600–700 °C. both water and carbon dioxide are released and the corre-

The MS study was done in vacuum, thus the results indi- cate, that $CO_2$ (possibly in form of carbonates) was present in the aged paste before heating, absorbed from air on age- ing.

IR spectroscopy study

The IR spectra of the aged paste C-43-St at various tem- of calcium carbonates (nominal at 1420–1480 $cm^{-1}$ [12]) in peratures are shown in Fig. 7. IR study shows the presence on heating up to 400 °C, decreasing at 600 °C. Also the wa- the hydrated and aged paste both at room temperature and ter, portlandite and silicate peaks were found, changing their form and/or disappearing with temperature (Table 3). peaks C2-3 (1485 and 1411 $cm^{-1}$), which decrease at 600 °C, when only calcite remains, whereas aragonite and All three polymorphs of $CaCO_3$ are shown by the joint vaterite are not stable [12,14,15,17].

The portlandite peak CH (3644 $cm^{-1}$) decreases with tem- perature and disappears above 400 °C [11,12,18,19]. 860 $cm^{-1}$) are observed up to 400 °C and at 600 °C they The silicate vibrations Si1-4 (1126, 995, 962, and are partly transformed into C2S (Ca2SiO4 = larnite, belite) [11,12,15].

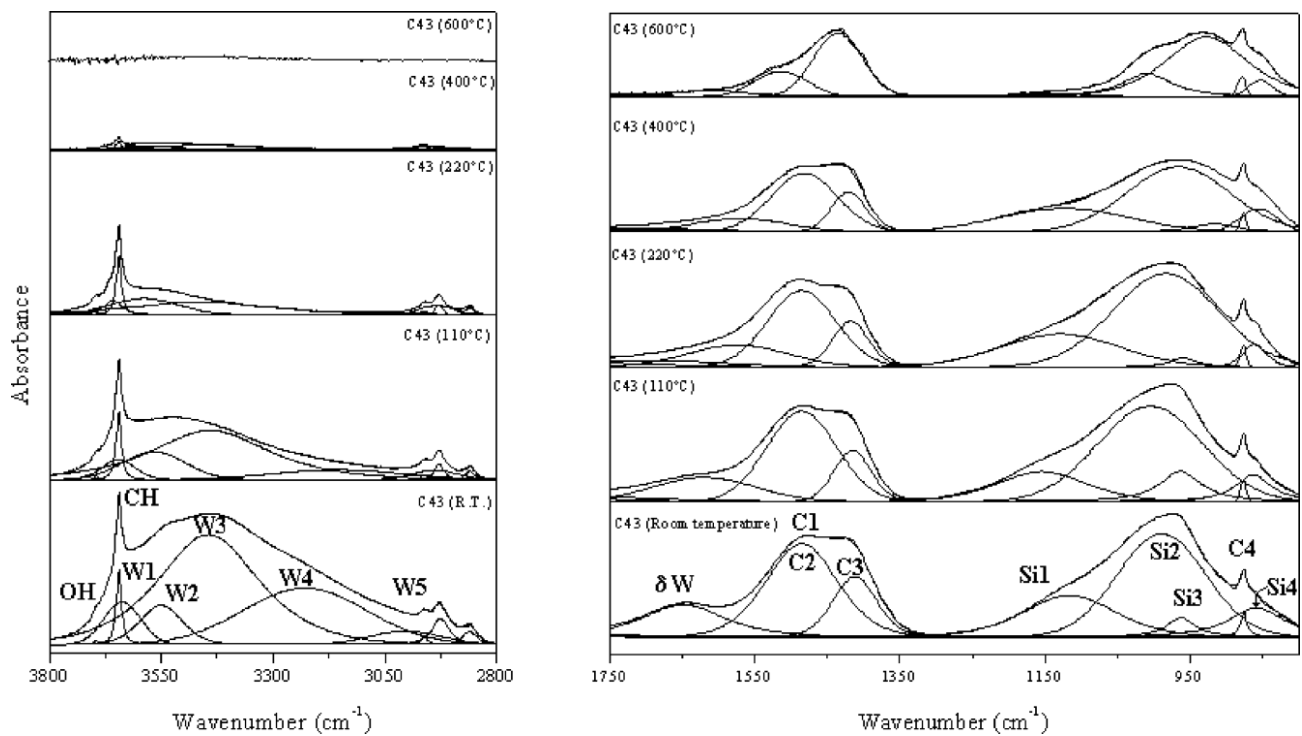

Fig. 7. IR spectroscopy of the aged paste C-43-St at various temperatures.

Table 3
FT-IR vibrations of C-43-St

| Symbol | Area | cm$^{-1}$ | Interpretation |
|---|---|---|---|
| OH | 0.11 | 3696 | Inner surface OH stretching [13] |
| CH | 3.35 | 3644 | Portlandite [11,12] |
| W1 | 9.72 | 3635 | Free water molecules ($v$(1) 3652 cm$^{-1}$) [14] |
| | | | Ettringite [11]. In smectites O–H stretching |
| | | | 3610–30 cm$^{-1}$. OH groups in water–water frequency with perpendicular polarization at |
| | | | H-bonds absorb at ~ 3400 cm$^{-1}$ |
| | | | Normal vibration frequencies of gaseous water $v$(1) 3657 cm−1, $v$2, 1595 cm−1, $v$3 3756 cm−1 [14] |
| W2 | 13.19 | 3549 | Water [11]; jennite (3565 cm−1 [15] |
| W3 | 73.54 | 3440 | The symmetric and antisymmetric stretching broad band centered at ∼3400 cm−1 (a) [14,12], vibrations ($v$1 and $v$3) of sorbed water gives a 3700 cm−1 (b) in free water; the reduction is due to formation of moderate H-bonds |
| W4 | 32.76 | 3236 | A shoulder near 3250 cm−1 = overtune of water bending vibrations near 1630 cm−1 [14] |
| W5 | 31.30 | 3037 | |
| OM1 | 0.29 | 2960 | Organic matter [16] 2950 cm−1 ($v_{as}$ CH3), ca. 2920 cm−1 ($v_{as}$ CH2), ca. 2850 cm−1 ($v_s$ CH2) |
| OM2 | 2.81 | 2925 | See above |
| OM3 | 1.10 | 2857 | See above |
| | 8.00 | 1881 | |
| C1 | 0.56 | 1794 | Calcite, aragonite, vaterite [15]. |
| δW | 22.99 | 1649 | Deformation vibrations $v$2 of adsorbed water molecules at 1630 cm−1 [14], characteristic of molecular water |
| | | | Bending vibrational band shifted from 1630 cm−1, indicates a greater restriction due to incorporation or association of water molecules into the cement matrix [12] |
| C2 | 51.23 | 1485 | Aragonite [14], vaterite; $v$(3) = 490 [14], 1473 [16], 1420–1480 cm−1 [12] |
| C3 | 19.94 | 1411 | Calcite [14], $v$(3) = 1407–1435 cm−1 |
| Si1 | 22.92 | 1126 | Si–O stretching vibrations 1120 and 1145 cm−1 in unhydrated cement [11]. Mollah et al. [12] interpret the 1126 cm−1 band as belonging to sulfate S–O stretching band |
| Si2 | 66.24 | 995 | Larnite [15], possibly HCO3− |
| Si3 | 9.59 | 962 | Possibly silicates [15] and HCO3−. The situation of the Si1-3 peaks indicates a high polymerization in the hydrated cement. The bands in dry cement are 925, 525 and 455 cm−1 [12] |
| C4 | 1.02 | 875 | $v$(2) calcite + aragonite, possibly larnite [15] |
| Si4 | 16.14 | 860 | 800 cm−1 are due to nesosilicates; absorption Four adsorption bands between 1000 and bands near 975 and 875 cm−1 are large and |
| | | | Fifth peak between 625 and 550 cm−1 [15] fairly broad; sharp peak on low frequency side. |
| | 2518 | >400 °C | Calcite [15] 2530–2500 cm−1 |
| | | 714 | Calcite [14] |
| | | 518 | Larnite [15] 521–523 cm−1 |

The stretching vibrations of water molecules ($v$1 + $v$3) show a broad band between 3000 and 3600 cm−1 (W1-5) and another one at 1649 cm−1 (δW) [11,14,15,20,21].

pear completely at 600 °C; they are due to organic matter [16]. HCO3− may be present (vibration at 1411 cm−1) [15]. The peaks OM1-3 decrease with temperature and disap-

The small peak OH may be interpreted as inner surface stretching vibrations [13].

Thus an equivocal evidence was obtained in IR study of the presence of calcium carbonates in the hydrated and aged paste both at room temperature and on heating up to 400 °C. The intensity of their peaks decreased at 600 °C.

3. Conclusions

- The assumptions made on estimation of cement hydration products by SH in the fresh paste proved correct.
- The sorbed water (EV), escaping at 110 °C from the fresh escaped at higher temperatures: partly up to 150 °C and partly between 500 and 700 °C. The second contribution paste, was bound on ageing with a higher energy and was much higher in the less compact paste than in the
standard one (11–13% in C-43-I, 2.5% in C-43-St).

- The joint water content of hydrates and of C-S-H gel (110–400 °C in SH, 150–450 °C in TG) increased on age- ing by 1–2% in the dense paste C-43-St and did not change
in the less compact one C-43-I. (O.M Formation)
- C-S-H gel transformed on heating above 600 °C into C2S and C3S.
- Portlandite decomposed at 450–470 °C and above 500 °C it transformed partly into calcite. This was measured in SH as sensitivity to carbonation.
- Portlandite content did not change on ageing. It was simi- lar in fresh paste, SH, and in the aged one, heated in argon,
TG. In the air atmosphere it became partly carbonated, which was accompanied by an increase in mass between 500 and 600 °C.
- Some form of carbonate was present in the aged paste The $CO_2$ evolution proceeded at 680–720 °C (TG) and at 605 °C in vacuum (MS), resulting in the XRD peaks of at low temperatures, transforming on heating into calcite.
CaO.
- Carbon dioxide and/or carbonate ions to form carbonates, were sorbed during ageing and were present in the aged
paste in some form undetectable by XRD (amorphous or crypto-crystalline).
- Sensitivity to carbonation [ΔM(600–800 °C) in SH] in- creased highly with ageing [ΔM(600–750 °C) in TG], es- pecially in the less compact paste. In the standard one it was similar whether heated in air or in argon. Thus a less compact structure facilitates the sorption of water (and its strong bonding) and of $CO_2$.

**Acknowledgements**

This research has been supported by research project MAT 2002-03774 from the Spanish Ministry of Science and Tech- nology and Research Group FQM-187 of the Junta de An- daluc´ıa.